\documentstyle[aps,prl,floats,epsfig]{revtex}
\begin{document}
\draft \twocolumn[\hsize\textwidth\columnwidth\hsize\csname
@twocolumnfalse\endcsname \title{Superflow-Stabilized Nonlinear NMR in
  Rotating $^3$He-B} \author{V.V. Dmitriev$^a$, V.B. Eltsov$^{ab}$, M.
  Krusius$^b$, J.J. Ruohio$^b$, and G.E.Volovik$^{bc}$} \address{
  $^a$Kapitza Institute for Physical Problems, 117334 Moscow,
  Russia\\
  $^b$Low Temperature Laboratory, Helsinki
  University of Technology, Box 2200, FIN-02015 HUT, Finland\\
  $^c$ L.D. Landau Institute for Theoretical Physics, 117334 Moscow,
  Russia }

\date{\today} \maketitle

\begin{abstract}
  Nonlinear spin precession has been observed in $^3$He-B in large
  counterflow of the normal and superfluid fractions. The new
  precessing state is stabilized at high rf excitation level and
  displays frequency-locked precession over a large range of frequency
  shifts, with the magnetization at its equilibrium value. Comparison
  to analytical and numerical calculation indicates that in this state
  the orbital angular momentum $\vec L$ of the Cooper pairs is
  oriented transverse to the external magnetic field in a
  ``non-Leggett'' configuration with broken spin-orbit coupling. The
  resonance shift depends on the tipping angle $\theta$ of the
  magnetization as $\omega -\omega_L = (\Omega_B^2/2\omega_L)(\cos
  \theta - 1/5)$. The phase diagram of the precessing modes with
  arbitrary orientation of $\vec L$ is constructed.
\end{abstract}
 \bigskip
] 

The study of nonlinear NMR response in superfluid $^3$He started with
the discovery of the Brinkman--Smith (BS) mode in $^3$He-A
\cite{Smith-Brinkman1} and $^3$He-B \cite{Smith-Brinkman2} phases.
These became the classic examples of nonlinear spin resonance in
magnetically ordered superfluids. Their observation by Osheroff and
Corruccini \cite{OsheroffCorruccini} opened the road to the discovery
of new resonance states in $^3$He-B, such as space-coherent precession
within a Homogeneously Precessing Domain (HPD) \cite{Bunkov1} or the
newly found stable modes where the magnitude of the precessing
magnetization differs from its equilibrium value \cite{HMletter}. An
example of the latter is the family of half-magnetization modes (HM),
where the precessing magnetization equals one half of the equilibrium
value \cite{Kharadze}. All of these are stable dynamic order parameter
states and nonlinear solutions of the Leggett--Takagi spin dynamic
equations. In principle, such states are similar to Q-balls, the
coherent soliton-like nontopological states of relativistic quantum
field theories, whose frequency and stability are determined by the
conservation of the global charge, say, the baryonic charge
\cite{Qball}. In $^3$He-B spin dynamics the role of the global charge
is played by the projection of the spin $S_z$ in the direction of the
magnetic field $\vec H$, which determines, in part, the NMR frequency
shift.

In these resonance modes the distinguishing factor is the orientation
of the orbital angular momentum $\vec L$ of the Cooper pairs.  In the
BS and HPD modes, $\vec L$ is oriented along the applied magnetic
field $\vec H$ via the spin-orbit (dipole) coupling, i.e. $\vec L
\parallel \vec H$. In contrast, the HM modes form with $\vec L$
oriented spontaneously perpendicular to $\vec H$. So far, in the
nonlinear regime the orientation of $\vec L$ has not been controlled
by external means. Here we use vortex-free counterflow of the normal
and superfluid components in a rotating container, to orient $\vec L$
along the flow direction. At high rotation velocity $\Omega$, the
orienting effect on $\vec L$ from the flow far exceeds that from the
dipole coupling. If the external magnetic field is oriented along the
rotation axis ($\vec H \parallel \vec \Omega$), one can then study the
unusual situation, when $\vec L \perp \vec H$. By measuring the
tipping angle of the precessing magnetization $\vec M$ as a function
of the applied frequency shift, we identify a new nonlinear resonance
mode which greatly differs from the classic case of $\vec L \parallel
\vec H$. In the linear regime the condition $\vec L \perp \vec H$ has
been realized in earlier measurements in the parallel-plate geometry
or in the presence of counterflow \cite{Hakonen1}.

{\it Experiment.}---Our cw NMR setup in the rotating nuclear
demagnetization cryostat has been described in
Refs.~\cite{HMletter,Ruutu1}. The $^3$He NMR sample is contained in a
quartz glass cylinder with a radius $R=2.5$ mm.  The measurements are
performed at fixed frequency $\omega_{\rm rf}/2\pi= 688$ kHz, using a
linear field sweep centered around a Larmor field value of $H=21.2$
mT, with a homogeneity $\Delta H/H = 2\cdot 10^{-4}$ over the sample
volume.  The signal is read with a lock-in amplifier, such that the
component in phase with the excitation field $H_{\rm rf}$ is called
dispersion $(\propto M_x = M_\perp \cos{ \phi})$ and the out-of-phase
component absorption $(\propto M_y = M_\perp \sin{ \phi})$. The
measuring range comprises counterflow velocities $\Omega R \leq 6.5$
mm/s, rf fields up to 0.03\,Oe, temperatures (0.7---1)\,T$_c$, and
pressures 0---12\,bar.

\begin{figure}[!!!tb]
  \centerline{\epsfig{file=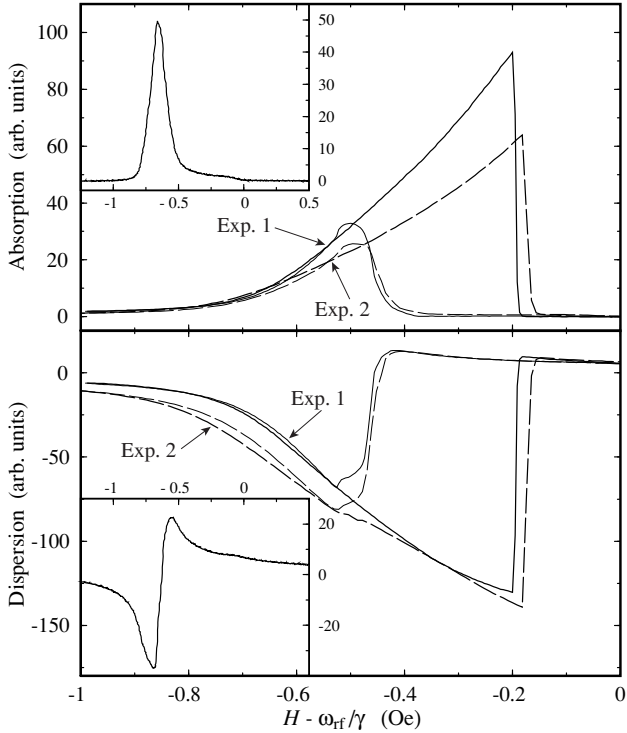,width=0.95\linewidth}}
  \medskip
  \caption{NMR response  while the applied magnetic field $H$ 
    is swept at constant rate $\dot H$ during continuous rf
    irradiation at fixed frequency $\omega_{\rm rf}$ ($\Omega = 2.5$
    rad/s, $P=0\,$bar, and $T\approx 0.88\, T_c$). Up sweeps are
    marked with thicker lines and down sweeps with thinner lines.
    Solid lines correspond to $H_{\rm rf} = 0.02\,$Oe and $\dot H =
    0.017\,$Oe/s (Exp. 1), dashed lines to $H_{\rm rf}=0.027\,$Oe and
    $\dot H=0.034\,$Oe/s (Exp. 2). Note that at higher excitation
    lower absorption is measured for compensating the relaxation.
    {\it (Insets)} Line shapes of the signal in the linear regime at
    low exitation level ($\Omega = 1.3$ rad/s, $P=2.0$ bar, $T= 0.90\;
    T_c$, and $H_{\rm rf} = 0.003$ Oe).}
  \label{NMRresponse}
\end{figure}

At low excitation amplitude ($H_{\rm rf} \sim 0.003\,$Oe), when the
tipping angle is a few degrees, the NMR response of $^3$He-B exhibits
linear behaviour: The line shapes of the absorption and dispersion
signals are independent of sweep direction, and the signal amplitudes
increase linearly with excitation level. In vortex-free counterflow at
sufficiently high velocity ($\Omega \gtrsim 0.5$ rad/s) the NMR
absorption maximum is shifted from the Larmor value~\cite{Hakonen1},
as shown by the NMR spectrum in the inset of Fig.~\ref{NMRresponse}.

\begin{figure}[!!!tb]
  \centerline{\epsfig{file=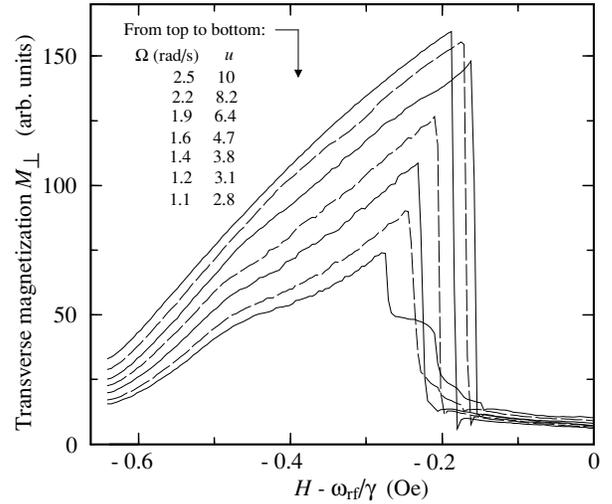,width=0.9\linewidth}}
  \medskip
  \caption{Transverse magnetization component
    $M_\perp$ as a function of field $H$, at different counterflow
    velocities in vortex-free rotation (listed in terms of $\Omega$
    and the dimensionless variable $u$, calculated at the outer
    perimeter of the sample cylinder from
    Eq.~(\protect\ref{Normalization})).  Only upward sweeps of the
    field $H$ are shown, down sweeps depend less on $\Omega$ and are
    similar to those in Fig.~\protect\ref{NMRresponse}. The
    temperature increases slowly during the measurement from
    $0.88\,T_c$ at the highest velocity to $0.92\,T_c$ at the lowest
    velocity. The normal state magnetization $\chi_{\rm N} H$
    corresponds approximately to 200 units on the vertical scale
    ($P=0\,$bar, $H_{\rm rf}=0.027\,$Oe, and $\dot H=0.017\,$Oe/s).}
  \label{veldep}
\end{figure}

At high excitation levels the absorption and dispersion signal
amplitudes increase faster than the rf field, and become highly
asymmetric (Fig.~\ref{NMRresponse}). The nonlinear behavior becomes
most pronounced while scanning the field in the upward direction
towards the Larmor value until finally an abrupt jump appears from the
precession at large tipping angle $\theta$ to the linear NMR regime
with small $\theta$. If the excitation amplitude is increased, the
jump usually moves to higher field. The field sweep in the opposite
direction has significantly different shape and the regime of large
tipping angles is not entered. The magnitude of the counterflow plays
a crucial role (Fig.~\ref{veldep}): With decreasing $\Omega$ both the
maximum tipping angle and the range of frequency shifts quickly
decrease.  This means that the new state appears only at high
counterflow velocities above the textural transition in which $\vec L$
is deflected into the plane transverse to $\vec \Omega$ in a
significant part of the cross section of the sample cylinder
\cite{FlowTextures}.

We interprete these observations in the following manner. For fixed
orientations of $\vec S$ and $\vec L$, the resonance frequency shift
is determined by the dipole interaction. During the field sweep these
vectors deflect from there equilibrium positions and the frequency of
the resonance absorption changes. If the sweep is performed in a
suitable direction, it may become possible to create a state where the
resonance frequency stays locked to the external excitation frequency.
In this case the increasing deflection of $\vec S$ causes the
transverse magnetization $M_\perp$ to increase, which is observed in
the experiment during a sweep towards the Larmor value as an increase
in the dispersion and absorption signals. The tipping angle increases
continuously with the field sweep as long as the rf pumping is
sufficient to compensate for relaxation, which increases as the
deviation of $\vec S$ from its equilibrium orientation increases.
Finally the mode collapses, in a first order transition between two
different dynamic order parameter states. This behaviour resembles
that of an anharmonic oscillator in forced oscillation. The remarkable
feature of $^3$He-B is that the rigidity from the order parameter
coherence makes the superfluid to behave like a single oscillator.

{\it Classification of modes.}---An analytic description of spin
precession with arbitrary orientations of $\vec S$ and $\vec L$ can be
constructed, if we neglect magnetic relaxation and the interaction
with the excitation field. The orientation of the orbital momentum,
below denoted by the unit vector $\hat l=-\vec L/|\vec L|$, is fixed
by the balance between its interactions with the counterflow and with
the precessing spins via the dipole coupling. The former is written as
\begin{equation}
F_{\rm cf}=-{1\over 2}\rho_a((\vec v_s -\vec
v_n)\cdot 
\hat l)^2~~.  
\label{CounterflowTerm}
\end{equation}
Here $\rho_a = \rho_{s \perp} - \rho_{s \parallel}$ is the superfluid
density anisotropy, caused by the magnetic field \cite{Korhonen2}. The
relative magnitude of the counterflow and dipole energies is
conveniently expressed in terms of a dimensionless velocity:
\begin{equation}
u={{15\rho_a \gamma^2 (\vec v_s -\vec v_n)^2}\over
{4\chi\Omega_B^2}} \; , 
\label{Normalization}
\end{equation}
where $\Omega_B (T,P)$ is the characteristic $^3$He-B frequency and a
measure of the dipole energy. Spin precession is simplified in the
high-field limit, when the dipole term is small compared to Zeeman
energy, $-\vec S\cdot \gamma\vec H$, and can be considered as a
perturbation. In zero order perturbation theory one has precession
with the Larmor frequency, $\omega_L=\gamma H$. A first order
correction gives the frequency shift from the precessing frequency
$\omega$: $ \omega -\omega_L=-\partial F_D/\partial S_z$, where $F_D$
is the dipole energy, averaged over the period of the precession.
Here we consider only the case when the precessing spin has its
equilibrium magnitude $S=\chi_{\rm B} H/\gamma$. For arbitrary
orientation of the orbital momentum, the time-averaged dipole energy
$F_D$ can be written as~\cite{KorhonenVolovik}
\begin{eqnarray}
\nonumber F_D=\frac{2}{15}\frac{\chi}{\gamma^2}\Omega_B^2[(s_z l_z-
{1\over 2}+{1\over 2}\cos\Phi(1+s_z)(1+l_z))^2 +\\{1\over
8}(1-s_z)^2(1-l_z)^2 +(1-s_z^2)(1-l_z^2)(1+\cos\Phi)]~. 
  \label{DipoleEnergyAverage}
\end{eqnarray}
The notations are: $s_z=S_z/|{\vec S}|=\cos{\theta}$, $l_z$ is the
projection of the orbital momentum $\hat l$ on the direction of the
magnetic field $\vec H$, and the angle $\Phi$ is a soft variable
related to the $^3$He-B order parameter.  The energy $F_D$ is also to
be stationary with respect to $\Phi$: $ \partial F_D/\partial \Phi
=0$.

Thus by varying the dipole energy with respect to the spin $S_z$ (the
analogue of the global charge) one obtains the precessing modes as a
function of the frequency shift and of the other global charge $L_z$,
which is kept fixed because of orbital viscosity.  Omitting the
limiting cases $\theta=0$ and $\theta=\pi$, we get three modes of
precession
\begin{eqnarray}
  1)&~~~~& \cos\Phi= - {(1 -2 l_z)
    (1 -2 s_z)\over (1 + s_z)(1 + l_z)},\nonumber\\
  &&\hskip 6em
  s_z ={3-18 l_z +15 l_z^2+ 4w \over 15 (1-l_z)^2},\\
  2)&~~~~& \cos\Phi=1,~~~~
  s_z=-{-1 + 4 l_z +5 l_z^2 + 4w\over -13 + 10 l_z
    +35 l_z^2 },\\
  3)&~~~~& \cos\Phi=-1,~~
  s_z=-\frac{3(1 - l_z^2) + 4w}{3(1- l_z)^2}.
\end{eqnarray}
Here $w$ is the dimensionless frequency shift:
\begin{equation}
w={{15\omega_L(\omega - \omega_L)}\over {2\Omega_B^2}} \; .
\label{FreqShift}
\end{equation}

\begin{figure}[!!!tb]
  \centerline{\epsfig{file=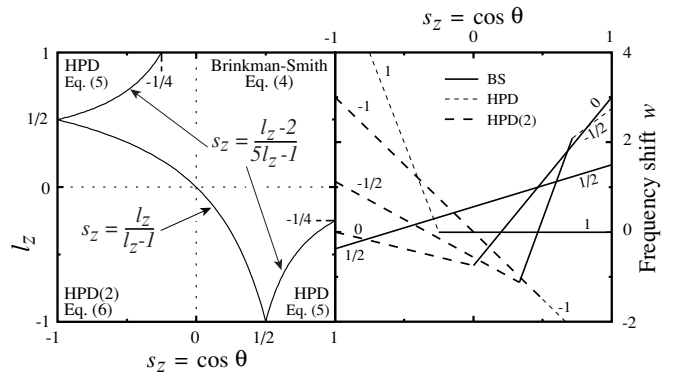,width=\linewidth}}
  \medskip
  \caption{{\it (Left)} Phase diagram of the precessing states for
    arbitrary orientations of $\vec L$ and $\vec S$. {\it (Right)}
    Dependence of the frequency shift $w$ on the tipping angle
    $\theta$ for a few fixed values of $\hat l_z$ (marked for each
    line at its end point).}
  \label{diagram}
\end{figure}

In Fig.~\ref{diagram} the $l_z$---$s_z$ phase diagram is shown with
the stable regions for each of the three modes. In the absence of
counterflow, $u=0$, the dipole coupling orients the orbital momentum
$\hat l$ along the magnetic field $\vec H$ in modes 1 and 2 and
opposite to the field in mode 3. Then mode 1 becomes the BS state with
zero frequency shift in the range $-{1\over 4} <s_z<1$ and $l_z =1$.
Mode 2 reduces to the HPD state with $-1<s_z<-{1\over 4}$, $l_z =1$,
and a frequency shift vs tipping angle dependence as $w=-2-8s_z$,
while mode 3 transforms to the so-called HPD(2) which has not been
seen experimentally \cite{BunkovVolovik}.  In the generalized phase
diagram of Fig.~\ref{diagram} with nonzero $u$, we retain these names
for the regions in which their respective $u=0$ modes lie. In large
counterflow linear NMR at small $\theta$ and nonlinear NMR at large
$\theta$ are located in Fig.~\ref{diagram} on the $l_z = 0$ axis and
belong to the same class of BS states. During an up sweep the
precession moves continuously from small to large $\theta$, until
ultimately magnetic relaxation causes an instability and a first order
transition takes $\theta$ back into the linear regime. This is similar
to a gas-liquid transition, where no symmetry break occurs.

{\it Limit of large counterflow.}---In rapid rotation $u$ becomes
large ($u\sim 15$) and the orbital momentum is rigidly forced into the
transverse plane over most of the cross section of the sample
cylinder: $ l_z=0$. The HPD state does not exist in this limit, while
the BS and HPD(2) modes have the frequency shifts
\begin{eqnarray}
&\displaystyle \omega   -\omega_L=
{\Omega_B^2\over 2\omega_L}\left(\cos \theta
-{1\over 5}\right)~,~~~~~s_z>0~.&
\label{FrequencyShiftDimensional}\\
&\displaystyle
\omega  -\omega_L=-{\Omega_B^2\over 10\omega_L}\left(1+\cos
\theta\right)~,~s_z<0~.&
\label{FrequencyShiftDimensionalHPD2}
\end{eqnarray}
In Eq.~(\ref{FrequencyShiftDimensional}) the linear regime of small
$\theta$ corresponds to $s_z \approx 1$ and $w=3$ or $\omega -
\omega_L = (2/5) \Omega_B^2/ \omega_L$.  In the nonlinear regime at
large $\theta$ the frequency shift has a positive slope as a function
of the longitudinal magnetization: $d \omega/d\cos\theta >0$.  This
suggests that in free precession in a pulsed NMR measurement
homogeneous precession should become unstable and break into domains.
In contrast, in continuous rf excitation the phase of the precession
is locked to that of $H_{\rm rf}$ and no instability occurs.  The
HPD(2) shift in Eq.~(\ref{FrequencyShiftDimensionalHPD2}) has a
negative slope, $d\omega/d\cos\theta <0$, and, since the conventional
HPD state is unstable in counterflow, HPD(2) is thus the only
inherently stable mode in large counterflow, with spontaneous
phase-coherence in free precession. However, the HPD(2) mode displays
large Leggett--Takagi relaxation because of its large deviation from
the Leggett configuration.  Presumably in the $T \rightarrow 0$ limit,
where relaxation vanishes, the HPD(2) mode might become observable.

\begin{figure}[!!!tb]
  \centerline{\epsfig{file=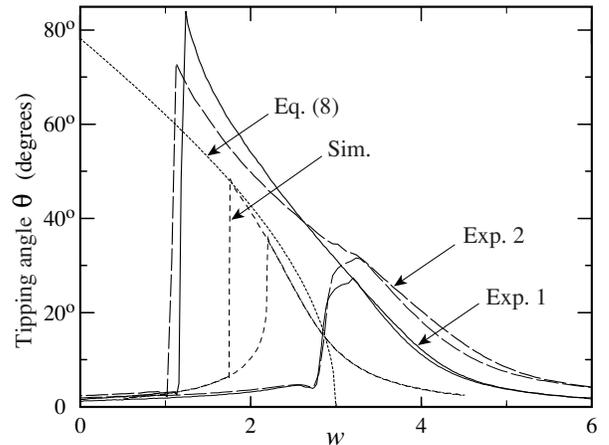,width=0.9\linewidth}}
  \medskip
  \caption{Tipping angle $\theta$ as a function of the
    normalized frequency shift $w$: Comparison of
    Eq.~(\protect\ref{FrequencyShiftDimensional}) to the measurements
    of Fig.~\protect\ref{NMRresponse}. A numerical solution is also
    shown with parameter values as in the experiment, except for
    $H_{\rm rf} = 0.014\,$Oe, $T= 0.95\,T_c$, $\dot H = 0.02\,$Oe/s,
    and spatially homogeneous counterflow at 0.70 cm/s.  Upward sweeps
    of the field are plotted with thick and downward sweeps with thin
    lines.}
  \label{comptheory}
\end{figure}

{\it Comparison with experiment.}---In Fig.~\ref{comptheory} we plot
the tipping angle $\theta$ from Eq.~(\ref{FrequencyShiftDimensional})
as a function of the frequency shift $w$, along with the two measured
NMR responses from Fig.~\ref{NMRresponse}. The frequency shift, at
which the new mode collapses in the experiment, is determined by the
relaxation processes. At the pressure of 12 bar, relaxation is less
and the new state is often stable during the upward sweep until above
the Larmor field value, i.e. to negative frequency shifts, in
agreement with the right panel of Fig.~\ref{diagram}. Overall, we
regard the agreement in Fig.~\ref{comptheory} as satisfactory, if we
allow for two experimental difficulties.

One uncertainty arises from assigning the proper value to temperature.
Temperature is measured in the linear NMR regime at low rf level,
using the fact that the counterflow absorption maximum is then
centered at $w=3$ (inset of Fig.~\ref{NMRresponse}). This frequency
shift determines the Leggett frequency $\Omega_B(T,P)$ in
Eq.~(\ref{FreqShift}) and, once calibrated \cite{Hakonen1}, can be
used as a thermometer with a sensitivity better than $0.001 \, T_c$.
In the NMR response at high rf level the only feature, which qualifies
for thermometry, is the maximum of absorption during the down sweep
(Fig.~\ref{NMRresponse}). Its location is not exactly $w=3$. In fact,
our numerical simulations suggest that it is $w < 3$. This means that
we ascribe a higher temperature and smaller $\Omega_B$ to our data
than actually would be the case, which explains why the measurements
are shifted to the right in Fig.~\ref{comptheory}.

Another factor is the heating of the sample by the absorbed rf power.
The sample cylinder is connected with a narrow channel to the
refrigerator \cite{Ruutu1}, to prevent vortices from leaking into the
NMR volume. The thermal resistance of the channel can lead to
unaccounted temperature rise and distortion of the line shape during
the field sweep. The heating is less important with a faster rate
$\dot H$ of sweep. All data in this paper were measured at large rates
so that the responses for up and down sweeps agree in their overlap
region at small tipping angles, indicating that the temperature is
approximately constant during the field cycling. The overall
uncertainty in temperature we estimate to about $\Delta T = +
0.02\,T_c$. This is sufficient to explain the difference between
measurement and Eq.~(\ref{FrequencyShiftDimensional}) in
Fig.~\ref{comptheory}.

{\it Numerical simulation.}---We have supplemented the analytic
description with direct numerical solution of the Leggett--Takagi
equations, by calculating the response of the spin-dynamic variables
in the time domain for the spatially homogenous case during continuous
rf excitation. Counterflow gives rise to the orientational energy
Eq.~(\ref{CounterflowTerm}) and to an additional torque, which
contributes to Leggett--Takagi relaxation: $\vec T=\delta F_{\rm
  cf}/\delta \vec\theta$, where $\delta \vec\theta$ is an
infinitesimal 3-dimensional rotation in spin space. The experimental
parameters $(T,P,H, \dot H, H_{\rm rf},\Omega)$ are adjusted to match
the experimental conditions.  A typical result is shown in
Fig.~\ref{comptheory}. It agrees surprisingly closely with
Eq.~(\ref{FrequencyShiftDimensional}), demonstrating that the effects
from relaxation and rf irradiation to the frequency shift are small.
It also reproduces the shape of the measured NMR response, showing
that the main difficulty in the comparison is the shift of the
measured data to a higher temperature. In general, the simulation
result is found to move closer to
Eq.~(\ref{FrequencyShiftDimensional}), when $T$, $H_{\rm rf}$, or
$\Omega$ are increased. These are all changes, which help to boost
either the value of $u$ or improve the compensation for relaxation,
and thus enhance the stability of the new mode towards larger fields
during the field sweep.

{\it In conclusion,} we have observed a highly nonlinear NMR response
of $^3$He-B when the direction of the orbital momentum is fixed by
large counterflow. This new state of precession has been identified as
a Brinkman--Smith mode in a general classification scheme of states
with fixed direction of $\vec L$. Other branches of the phase diagram
can perhaps be found in experiments where the direction of $\vec L$ is
fixed with solid walls or by tilting the magnetic field towards the
flow direction.


\vspace*{-4mm}
\begin{references}
  
  \vspace*{-12mm}
  
  
\bibitem{Smith-Brinkman1} W.F. Brinkman, H. Smith, Phys. Lett.{\bf
    51~A}, 449 (1975).
  
\bibitem{Smith-Brinkman2} W.F. Brinkman, H. Smith, Phys. Lett. {\bf
    53~A}, 43 (1975).
  
\bibitem{OsheroffCorruccini} D.D.  Osheroff and L.R.Corruccini, Phys.
  Lett. {\bf 51~A}, 447 (1975); in {\it Proc. 14th Intern. Conf. Low
    Temp. Phys. LT-14} Vol.1 (North-Holland, Amsterdam, 1975), p.
  100.
  
\bibitem{Bunkov1} See review by Yu.M. Bunkov, in {\it Prog. Low Temp.
    Phys.} Vol.  XIV (Elsevier Science, Amsterdam, 1995), p. 69.
  
\bibitem{HMletter} V.V. Dmitriev {\it et al.}, Phys. Rev. Lett. {\bf
    78}, 86 (1997).
  
\bibitem{Kharadze} G. Kharadze {\it et al.}, J. Low Temp: Phys. {\bf
    110}, 851 (1997).
  
\bibitem{Qball} S. Coleman, Nucl. Phys. {\bf B262}, 263 (1985); more
  recent eg. A. Kusenko {\it et al.}, Phys. Rev. Lett. {\bf 80}, 3185
  (1998).
  
\bibitem{Hakonen1} See eg. P.J. Hakonen {\it et al.}, J. Low Temp.
  Phys. {\bf 76}, 225 (1989), and references therein.
  
\bibitem{Ruutu1} V.M. Ruutu {\it et al.}, J. Low Temp. Phys. {\bf
    107}, 93 (1997).
  
\bibitem{FlowTextures} J.S. Korhonen {\it et al.}, Phys. Rev. Lett.
  {\bf 65}, 1211 (1990).
  
\bibitem{Korhonen2} J.S. Korhonen {\it et al.}, Phys. Rev. B{\bf 46},
  13983 (1992).
  
\bibitem{KorhonenVolovik} J.S.  Korhonen, G.E. Volovik, JETP Lett.
  {\bf 55}, 362 (1992). The time averaged dipole energy was first
  discussed by I.A. Fomin (J. Low Temp. Phys. {\bf 31}, 509 (1978)).
  
\bibitem{BunkovVolovik} Yu.M. Bunkov, G.E. Volovik, JETP {\bf 76},
  794, (1993).

\end{references}
\end{document}